\documentclass[3p,times,twocolumn]{elsarticle}

\usepackage{ecrc}
\usepackage{aas_macros}

\volume{00}

\firstpage{1}

\journalname{Nuclear and Particle Physics Proceedings}

\runauth{H. Kang}

\jid{nppp}

\jnltitlelogo{Nuclear and Particle Physics Proceedings}

\usepackage{amssymb}
\usepackage[figuresright]{rotating}

\begin{document}

\begin{frontmatter}



\dochead{}

\title{Particle Acceleration at Structure Formation Shocks}


\author{Hyesung Kang}

\address{Department of Earth Sciences, Pusan National University, Busan 46241, Republic of Korea}
\ead{hskang@pusan.ac.kr}

\begin{abstract}
Cosmological hydrodynamic simulations have demonstrated that shock waves
could be produced in the intergalactic medium by supersonic flow motions during the course of hierarchical
clustering of the large-scale-structure in the Universe. 
Similar to interplanetary shocks and supernova remnants (SNRs), these structure formation shocks can
accelerate cosmic ray (CR) protons and electrons via diffusive shock acceleration.
External accretion shocks, which form in the outermost surfaces of nonlinear structures, 
are as strong as SNR shocks and could be potential accelerations sites for high energy CR protons up to $10^{18}$~eV.
But it could be difficult to detect their signatures due to extremely low kinetic energy flux associated with those accretion shocks. 
On the other hand, radiative features of internal shocks in the hot intracluster medium have been identified 
as temperature and density discontinuities in X-ray observations and diffuse radio emission from accelerated CR electrons.
However, the non-detection of gamma-ray emission from galaxy clusters due to $\pi^0$ decay
still remains to be an outstanding problem.
\end{abstract}


\end{frontmatter}

% \linenumbers

\section{Introduction}

In \cite{hilass84} shocks in the intracluster medium (ICM) appeared 
as candidate acceleration sites for ultra-high-energy cosmic rays (CRs) in the so-called `Hillas diagram',
in which the maximum energy of CR nuclei achievable by a cosmic accelerator was estimated from the confinement condition: 
\begin{equation}
E_{\rm max}({\rm ZeV})\sim z \cdot \beta_a\cdot  B_{\mu G}\cdot L_{\rm Mpc},
\end{equation}
where $E_{\rm max}$ is given in units of $10^{21}$~eV, $z$ is the charge of CR nuclei, and $\beta_a= v_a/c$, $B_{\mu G}$, and $L_{\rm Mpc}$ are the characteristic speed, the magnetic field strength in units of microgauss, 
and the size in units of Mpc of the accelerator, respectively.
For shocks associated with galaxy clusters with $\beta_a\sim 0.01$, $B_{\mu G}\sim 1$,  $L_{\rm Mpc}\sim 1$,
CR protons could be accelerated up to $\sim 10^{19}$~eV.

\cite{norman95} first suggested that cosmic shocks induced by the structure formation can accelerate
CR protons up to $10^{19.5}$~eV via diffusive shock acceleration (DSA). 
Independently and more or less simultaneously, \cite{kang96} showed,
using cosmological hydrodynamic simulations, 
that accretion shocks around galaxy clusters have
$v_s\sim 3\times 10^3~{\rm km~s^{-1}}$, and suggested that, for the Bohm diffusion with microgauss magnetic fields,
the maximum energy of protons achieved via DSA by cluster accretion shocks
is limited to $\sim 60$~EeV ($\tau_{\rm acc}=\tau_{\rm pion}$), due to the energy loss via photo-pion interactions with the cosmic background radiation (see Figure 1).
Adopting simple models for magnetic field strength and DSA, 
and an analytic relation between the cluster temperature and the spherical accretion shock,
\cite{kang97} showed that the CR protons from a cosmological ensemble of cluster 
accretion shocks could make a significant contribution to the observed CR flux near $10^{19}$~eV.

\begin{figure}[t!]
%\centering
\vskip -0.0cm
\hskip -0.0cm
\includegraphics[width=70mm]{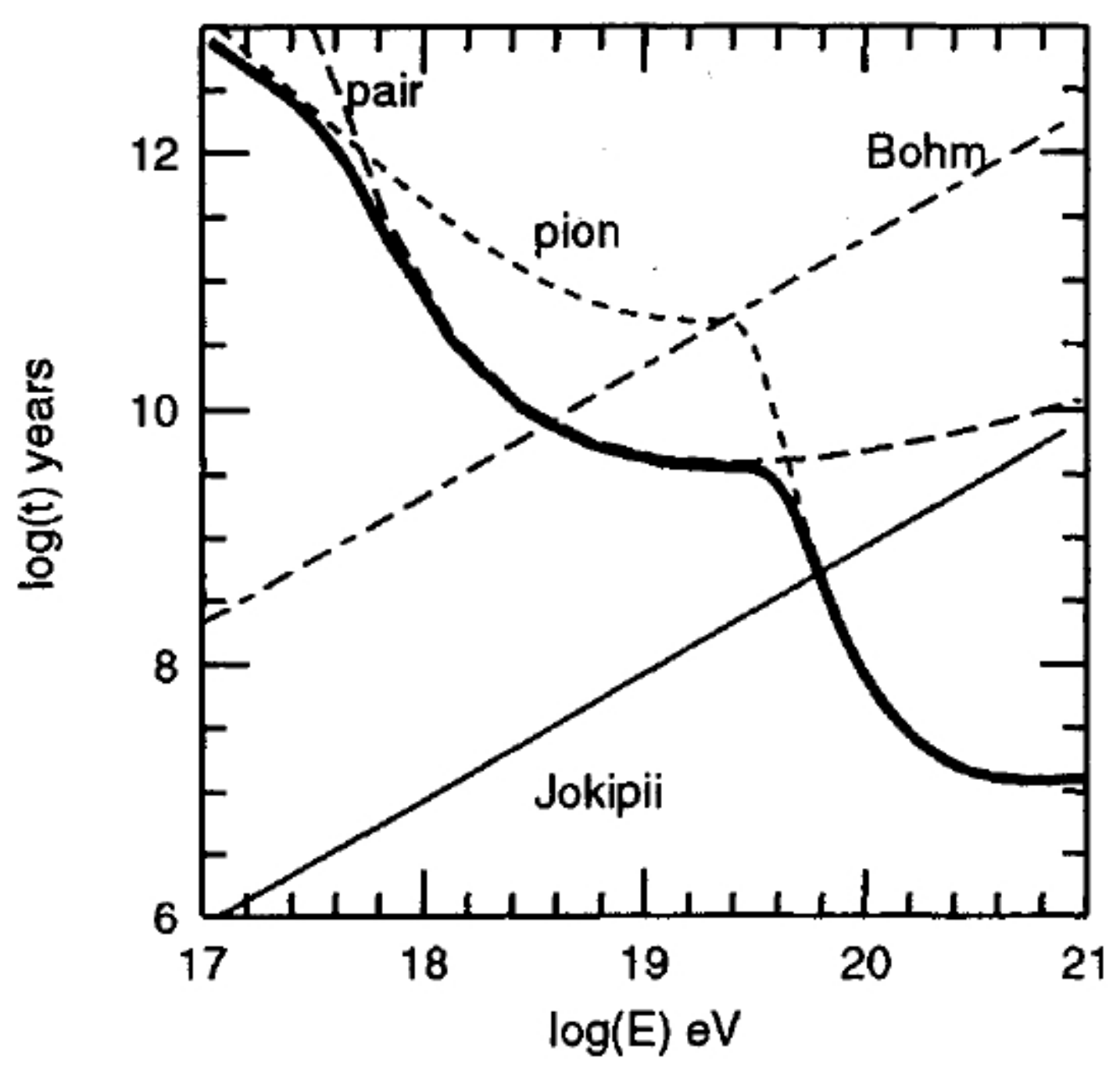}
\vskip -0.0 cm
\caption{Energy loss time scales for CR protons due to pair-production ($\tau_{\rm pair}$, thick dashed line) 
and pion-production ($\tau_{\rm pion}$, thin dashed) 
on the cosmic background radiation. The thick solid line represents the time scale due to the sum of the two loss processes. 
Shock acceleration time scales for Bohm ($\tau_{\rm Bohm}$, dot-dashed) and Jokipii ($\tau_{\rm Jokippi}$, thin solid) diffusion
at the shock with $u_s=10^3~{\rm km~s^{-1}}$ and $B=1~\mu {\rm G}$  \cite{kang97}.
}
\end{figure}

Observational evidence for the electron acceleration by a cluster accretion shock was first suggested by \cite{ensslin98}
who proposed that diffuse radio relics detected in the outskirts of several clusters could be diffuse synchrotron emission
from fossil electrons re-energized by accretion shocks.
Since the discovery of a shock in the Bullet cluster (1E 0657-56) \cite{markevitch02}, about a dozen of shocks have been detected
as sharp discontinuities in X-ray temperature or surface brightness in mainly merging clusters \cite{markevitch07,russell10}.
Moreover, giant radio relics such as the Sausage relic in CIZA J2242.8+5301 and the Toothbrush relic in 1RXS J0603.3+4214 are thought to result from merger-driven shocks,
since most of observed properties can be explained by synchrotron emission from shock-accelerated electrons cooling behind the shock
\cite{vanweeren10,vanweeren12}. So the presence of cosmic shocks in the ICM within a few Mpc from the cluster center has been established, although shocks in lower density filaments await to be detected by future observational facilities \cite{bruggen12,brunetti14}.  

In this contribution, we review the properties of structure formation shocks, 
the physical processes involved in the acceleration of CR ions and electrons at collisionless shocks, 
and observational signatures of shocks and nonthermal particles in the ICM.

\section{Properties of Structure Formation Shocks}

The properties and energetics of cosmological shocks have been studied extensively,
using numerical simulations for the large-scale-structure (LSS) formation
\cite[e.g.,][]{miniati00,ryu03,kang07,skillman08,hoeft08,vazza09}.
The average spatial frequency between shock surfaces is 
$\sim1~ {\rm Mpc}^{-1}$ inside nonlinear structures of clusters, filaments, and sheets.
These shocks can be classified mainly into two categories:
(1) external accretions shocks with the Mach number, $3\lesssim M_s \lesssim 100$, 
that form around the outermost surfaces of nonlinear structures, 
and (2) internal shocks mostly with $M_s \lesssim 5$ that form in the hot ICM inside nonlinear structures \cite{ryu03}.
In Figure 2, external accretion shocks encompassing the cluster coincide with the region with sharp temperature discontinuities, 
indicating high Mach number shocks. On the other hand, weak internal shocks within a few Mpc from the cluster center 
are associated with mild temperature variations.
The presence of internal shocks has been confirmed in many merging clusters,
while radiative signature of external accretion shocks have not been detected so far due to very low surface brightness.

\begin{figure}[t!]
%\centering
\vskip -0.7cm
\hskip -0.0cm
\includegraphics[width=80mm]{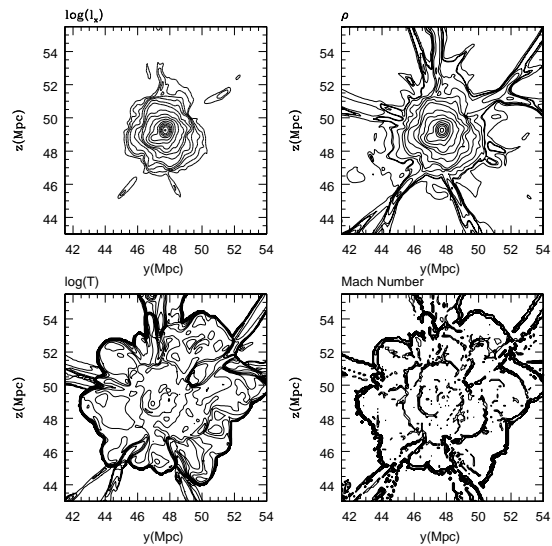}
\vskip -2.0cm
\caption{Two-dimensional slice showing X-ray emissivity, gas density, temperature, and shock locations around a galaxy cluster in a structure formation simulation. Strong external accretion shocks form in the outer surfaces of the cluster, 
while weak internal shocks reside inside the virialized central region \cite{ryu03}.
}
\end{figure}

Weak internal shocks with $2 \lesssim M_s\lesssim 3$ have high kinetic energy flux and are responsible for most of the shock energy
dissipation into heat and nonthermal components of the ICM such as CRs, magnetic fields, and turbulence.
By adopting a DSA model of CR proton acceleration, \cite{ryu03} predicted
that the ratio of the CR proton to gas thermal energies dissipated at
all cosmological shocks through the history of the Universe could be substantial, perhaps up to $50\%$.
However, this estimate has to be revised to significantly lower values as we will discuss in Section 5.

\section{Turbulence and Magnetic Fields in Large-Scale-Structure}
\label{}
Magnetic field is one of the key elements that govern the plasma processes at collisionless shocks and radiative signatures
of accelerated particles.
The intergalactic space is observed to be permeated with magnetic fields and
filled with turbulence and CRs, similar to the interstellar medium within our Galaxy
\cite{kang07,ryu08,dolag08,bruggen12,brunetti14}.
Analysis of the rotation measure data for Abell clusters indicates that the mean magnetic field strength ranges up to 
several $\mu{\rm G}$ in the ICM \cite{clarke01,carilli02}.

Using hydrodynamic and mageneto-hydrodynamic (MHD) simulations for the structure formation,
it has been suggested that turbulence could be produced in the ICM
by cascade of the vorticity generated behind cosmological shocks or by merger-driven flow motions,
and that the intergalactic magnetic fields could be amplified via turbulence dynamo 
\cite{ryu08,vazza09turb,cho14, miniati16}.
The seed fields might have been injected into the ICM via galactic winds and AGN jets or originate from some primordial processes
\cite{dolag08,bruggen12,brunetti14}.
This turbulence dynamo scenario typically predicts that the energy budget among different components in the ICM could be
$E_{\rm turb} \sim 0.1 E_{\rm th}$ and $E_{\rm B} \sim 0.01 E_{\rm th}$, where $E_{\rm th}$ is the thermal energy density  \cite{ryu08}.
As shown in Figure 3, 
the volume-averaged magnetic field strength ranges
$0.1-1~\mu {\rm G}$ in the ICM ($T>10^7$~K) and $0.01-0.1~\mu {\rm G}$ in filaments ($10^5<T<10^7$~K), 
which seems to be consistent with observations \cite{carilli02,clarke01}.
In the peripheral regions $\sim 5$~Mpc away from the cluster center where external accretions are expected to form,
the magnetic field strength should be similar to that of filaments, i.e., $\sim 0.01-0.1~\mu {\rm G}$.
The magnetic fields should be much weaker in sheet-like structures and voids,
but neither theoretical nor observational estimates are well defined in such low density regions.

\begin{figure}[t!]
%\centering
\vskip -3.0cm
\hskip -0.7cm
\includegraphics[width=90mm]{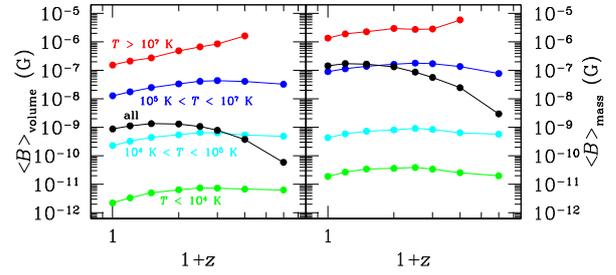}
\vskip -4.0cm
\caption{Magnetic field amplification based on turbulence dynamo in a structure formations simulation. 
Volume-averaged (left) and mass-averaged (right) magnetic field
strength as a function of redshift $z$ for the intergalactic medium in four temperature
ranges, $T > 10^7$~K (red, ICM), $T = 10^5 - 10^7$~K (blue, WHIM), $T = 10^4 - 10^5$~K (cyan), and
$T < 10^4$~K (green), and for all (black) the gas \cite{ryu08}.
}
\end{figure}

Relativistic protons and electrons with the same rigidity ($R=pc/ze$) are accelerated in the same way in DSA regime.
But for the particle injection to the DSA process, the obliquity angle, $\Theta_{\rm Bn}$, becomes an important factor.
At quasi-parallel shocks ($\Theta_{\rm Bn}\lesssim 45^{\circ}$), where the magnetic field direction is roughly 
parallel to the flow velocity,
MHD waves are self-generated due to streaming of CR protons upstream of the shock,
and protons are injected/accelerated efficiently to high energies via DSA \cite{bell04,caprioli14,caprioli14B}.
At quasi-perpendicular shocks ($\Theta_{\rm Bn}\gtrsim 45^{\circ}$), on the other hand, 
electrons tend to be reflected at the shock front and accelerated via shock drift acceleration (SDA)
and may further go through the Fermi I acceleration process, if they are scattered by plasma waves excited in the preshock region \cite{guo14}.

In addition to $\Theta_{\rm Bn}$, excitation of MHD/kinetic waves by plasma instabilities 
and wave-particle interactions at collisionless shocks
depend on the shock parameters such as the plasma beta, $\beta_p= P_{\rm gas}/P_{\rm B}$, and the Alfv\'en Mach number, 
$M_A\approx \sqrt{\beta_p} M_s$.
For the internal ICM shocks, $\beta_p\sim 50$, $M_s\lesssim 3$, and $M_A\lesssim 20$.
So they are super-critical, i.e., $M_A> M_{\rm crit}$, where the critical Mach Number is $M_{\rm cirt} \sim 1-1.5$ for high beta plasma,
and some ions are reflected specularly at the shock ramp, independent of the obliquity angle \cite{treumann09}.
In the foreshock region, some of incoming ions and electrons are reflected upstream,
and the drift between incoming and reflected particles may excite plasma waves via various micro-instabilities,
depending on the shock parameters.
For low beta plasma ($\beta_p\lesssim 1$), at high $M_A$ quasi-perpendicular shocks ($M_A\gtrsim \sqrt{\beta_p m_p/m_e}/2$)
the Buneman instability is known to excite electrostatic waves,
leading to the shock-surfing-acceleration of electrons in the shock foot \cite{amano09}.
For low $M_A$ quasi-perpendicular shocks ($M_A \lesssim \sqrt{m_p/m_e}/2$), on the other hand, 
the modified two stream instability could generate oblique whistler waves, which result in the pre-heating of thermal electrons
to a $\kappa$-like suprathermal distribution \cite{matsukiyo03}.

\section{Electron Acceleration at Cosmological Shocks}

\begin{figure}[t!]
%\centering
\vskip -1.0cm
\hskip -0.50cm
\includegraphics[width=90mm]{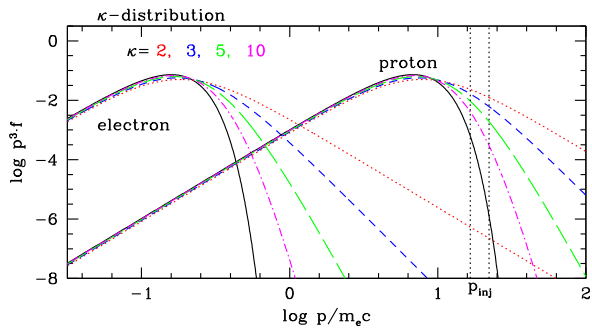}
\vskip -6.0cm
\caption{Momentum distribution, $p^3f(p)$, of electrons and protons for the gas with $kT\approx4.3$~keV
in the case of the $\kappa$-distributions with $\kappa=$ 2, 3, 5, and 10.
The Maxwellian distributions are shown in black solid lines.
The vertical lines indicate the range of the injection momentum of $p_{\rm inj}=
(3.5-4)\ p_{\rm th,p}$
above which particles can be injected into the DSA process \cite{kang14}.}
\end{figure}

Plasma kinetic processes govern the preacceleration of electrons in the shock transition zone, 
which leads to the injection of CR electrons to the Fermi I process.
Figure 4 shows thermal and suprathermal distributions of electrons and protons for the gas with $kT\approx4.3$~keV. 
The particle momentum should be greater than a few times the postshock thermal proton momentum ($p_{\rm th,p}$)
to cross the shock transition. 
So thermal electrons with $p_{\rm th,e}= p_{\rm th,p}\sqrt{m_e/m_p}$ need to be pre-accelerated to the injection momentum, 
$p_{\rm inj}\sim 3.5 p_{\rm th,p}$,
before they can start participating to the full DSA process \cite{malkov01}. 
Such injection from the thermal Maxwellian pool is expected to be very inefficient, especially
at low Mach number shocks, and depend very sensitively on the shock Mach number. 
But if there are suprathermal electrons with the $\kappa$-like power-law tail, instead of the Maxwellian distribution,
the injection and acceleration of electrons can be enhanced greatly even at weak cluster shocks \cite{kang14}.
As illustrated in Figure 4, the particle injection flux at $p_{\rm inj}$ is larger for a $\kappa$-distribution with
a smaller value of $\kappa$.
So the development of a $\kappa$-like suprathermal distribution is critical in the electron acceleration via DSA.

In the case of low $M_A$ quasi-perpendicular shocks in the high beta ICM plasma, 
some incoming electrons are mirror reflected at the shock ramp and gain energy via multiple cycles of SDA,
while protons can go through a few SDA cycles with only minimal energy gains \cite{guo14}.
In the foreshock of such weak shocks, the electron firehose instability induces oblique magnetic waves, which in turn provide efficient scattering
necessary to energize the thermal electrons to suprathermal energies, leading to efficient injection to the DSA process.
This picture is consistent with the observational fact that the magnetic field obliquity is typically quasi-perpendicular 
at giant radio relics such as the Sausage relic \cite{vanweeren10}, and the double relic in the cluster PSZ1 G108.18 \cite{degasperin15}.

Radio relics are diffuse radio structures detected in the outskirts of merging galaxy clusters.
Their observed properties can be best understood by synchrotron emission from relativistic electrons
accelerated at merger-driven shocks:
elongated morphologies over $\sim 2$~Mpc, spectral aging across the relic width (behind the putative shock),
integrated radio spectra of a power-law form with gradual steepening above $\sim 2 $~GHz,
and high polarization levels \cite{vanweeren10,vanweeren12,bruggen12,feretti12}.

The sonic Mach number of a relic shock can be estimated from either radio or X-ray observations, using
the radio spectral index relation, $\alpha_{\rm sh} = (M_{\rm rad}^2+3)/2(M_{\rm rad}^2-1)$, or
the X-ray temperature jump condition, $T_2/T_1=(M_X^2+3)(5M_X^2-1)/16/M_X^2$, respectively.
In some radio relics, the two estimates are different, i.e., $M_{\rm X} < M_{\rm rad}$, indicating that the simple DSA origin of radio relics
might not explain the observed properties \cite{akamatsu13}. For example, $M_{\rm X}\approx 1.2-1.5$ and $M_{\rm rad}\approx 3.0$
for the Toothbrush relic \cite{vanweeren16}, 
while $M_{\rm X}\approx 2.7$ and $M_{\rm rad}\approx 4.6$ for the Sausage relic \cite{vanweeren10,akamatsu15}.
Such discrepancy could be explained by the two following scenarios based on DSA:
(1) injection-dominated model in which $M_s\approx M_{\rm rad}$ and $M_{\rm X}$ is under-estimated due to projection effects in X-ray observation \cite{hong15}, and
(2) reacceleration-dominated model in which preexisting electrons with a flat energy spectrum is reaccelerated by a weak shock with $M_s\approx M_{\rm X}$
\cite{kang12,kang16JKAS}.
Figure 5 illustrates that such two viable scenarios, albeit with different sets of model parameters,
could reproduce the observed surface brightness and spectral index profiles of the Toothbrush relic \cite{vanweeren16}.

\begin{figure}[t!]
%\centering
\vskip -1.5cm
\hskip -0.5cm
\includegraphics[width=120mm]{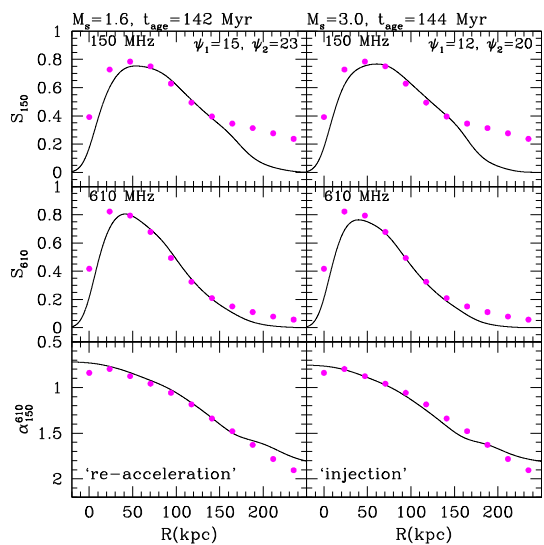}
\vskip -6.0cm
\caption{DSA modeling for the Toothbrush relic: reacceleration-dominated model with a $M_s\approx 1.6$ shock (left panels) 
and injection-dominated model with a $M_s=3.0$ shock 
(right panels). Radio flux density $S_{\nu}$ at 150~MHz (top) and at 610~MHz (middle),
and the spectral index $\alpha_{150}^{610}$ between the two frequencies (bottom) are
plotted as a function of the projected distance behind the shock (relic edge), $R(\rm kpc)$ \cite{kang16JKAS}.
The magenta dots are the observational data of the head portion of the Toothbrush relic \cite{vanweeren16}.}
\end{figure}

Using structure formation simulations, \cite{hong15} carried out mock observations of radio relic shocks 
detected in simulated clusters and 
showed that X-ray observations are inclined to detect weaker shocks due to projection effects, 
while radio observations tend to observe stronger shocks with flatter radio spectra.
This naturally supports the injection-dominated model, in which $M_{\rm X}$ tends to be smaller than $M_{\rm rad}$ for a given radio relic.

The ICM is thought to contain fossil relativistic electrons left over from tails and lobes of extinct AGNs.
Mildly relativistic electrons with $\gamma_e\lesssim 10^2$ survive for long periods of time,
since the cooling time scale of electrons in $B\sim 1~\mu {\rm G}$ is $t_{\rm rad}\approx 10^{10}~ {\rm yr}\cdot (10^2/\gamma_e)$.
They could provide seed electrons to the DSA process, which alleviates the low injection/acceleration efficiency problem 
at weak cluster shocks in the case of the injection-dominated model.
If we conjecture that radio relics form when the ICM shocks encounter fossil mildly relativistic electrons with 
$\gamma_e\lesssim 10^2$, 
then the model may explain why only about 10 \% of merging clusters contain radio relics \cite{kang16JKAS}.

The so-called infall shocks form in the cluster outskirts when the WHIM from adjacent filaments penetrates deeply into the ICM  \cite{hong14}.
They have relatively high Mach numbers ($M_s\gtrsim 3$) and large kinetic energy fluxes, 
so they could contribute to a significant fraction of CR production in clusters with actively infalling filaments.
So some radio relics with relatively flat radio spectra found in the cluster outskirts could be explained by 
these energetic infall shocks.

Although there still remain a few puzzles regarding the DSA origin of radio relics,
it is well received that shocks should be induced in merging galaxy clusters and radio relics could be
radiative signatures of relativistic electrons accelerated at those shocks \cite{bruggen12,brunetti14}.

\section{Proton Acceleration at Cosmological Shocks}

In the precursor of quasi-parallel shocks, CR protons streaming ahead of
the shock are known to excite both resonant and non-resonant waves and 
amplify the turbulent magnetic fields by orders of magnitude  \cite{lucek00,bell04}.
According to hybrid simulations by \cite{caprioli14}, the CR proton acceleration is efficient only for quasi-parallel shocks
with $\Theta_{\rm Bn}\lesssim 45^{\circ}$, 
and about $6-10\%$ of the postshock energy is transferred to CR proton energy for shocks with $M_A\sim 5-10$.
For quasi-perpendicular shocks, on the other hand,
protons go through only a few cycles of SDA before they advect downstream away from the shock.
So scattering waves are not self-generated in the preshock region, and thus the CR proton acceleration is very inefficient.
Note that for these simulations, $\beta_p\sim 1$ and $M_A\sim M_s$, so the quantitative estimates for the CR acceleration efficiency
may be different for the ICM shocks in high beta plasmas. 
Based on cosmological simulations with magnetic fields, the shock obliquity angle is expected to have the random orientation 
in the ICM and at surrounding accretions shocks \cite{ryu08}. 
Then the probability distribution function for the obliquity angle scales as $P(\Theta_{\rm Bn}) \propto \sin \Theta_{\rm Bn}$, 
so only $\sim 30\%$ of all cosmological shocks have the quasi-parallel configuration and accelerate CR protons \cite{vazza16CR}. 

Using cosmological hydrodynamic simulations, the $\gamma$-ray emission from galaxy clusters have been estimated 
by modeling the production of CR protons and electrons at structure formation shocks in several studies 
\cite[e.g.,][]{miniati01b,pfrommer07,pinzke10, vazza16CR}.
Inelastic collisions of shock-accelerated protons with thermal protons produce neutral pions, which decay into $\gamma$-ray photons
(hadronic origin) \cite{miniati01b}.
Inverse Compton upscattering of the cosmic background radiation by shock-accelerated primary CR electrons 
and by secondary CR electrons generated by decay of charged pions also provides $\gamma$-ray emission 
(leptonic origin) \cite{miniati01a}.
It has been shown that the hadronic $\gamma$-ray emission is expected to dominate over 
the leptonic contribution in the central ICM within the virial radius \cite[e.g.,][]{pinzke10}.

The key parameters in predicting the $\pi^0$ decay $\gamma$-ray emission are the CR proton acceleration efficiency,
$\eta(M_s)$, defined as the ratio of the CR energy flux to the shock kinetic energy flux,
and the volume-averaged ratio of the CR to thermal pressure in the ICM, 
$\langle X_{\rm CR}\rangle = \langle P_{\rm CR} \rangle / \langle P_{\rm th}\rangle $ \cite{kang07,kang13,pinzke10}.
Adopting the DSA efficiency model in which $\eta\approx 0.1$ for $M_s\approx 3$ given in \cite{ensslin07},
for example,
\cite{pinzke10} estimated that $\langle X_{\rm CR}\rangle \approx 0.02$ for Coma-like clusters.
In \cite{kang13}, in which a thermal-leakage injection model was implemented to DSA simulations,
the efficiency is estimated to be $\eta\approx 0.01-0.1$ for $M_s\approx 3-5$ shocks.
Note that in this DSA model the efficiency depends sensitively on the assumed injection model as well as $M_s$. 

\begin{figure*}[t]
%\centering
\vskip 0.0cm
\hskip +1.0cm
\includegraphics[width=140mm]{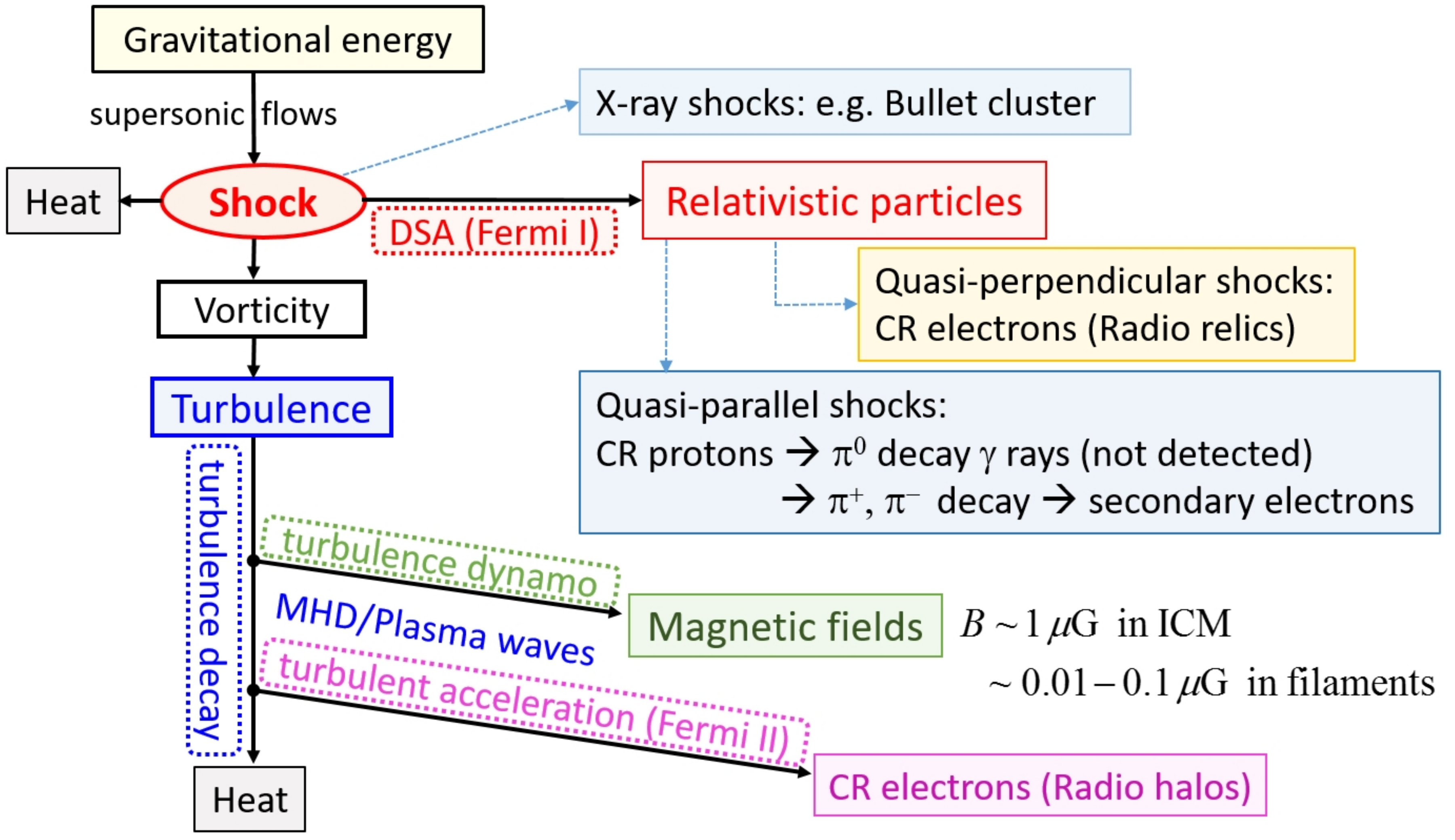}
\vskip +0.0cm
\caption{Physical processes and observational signatures expected to operate at structure formation shocks}
\end{figure*}

Recently, \cite{vazza16CR} tested several different prescriptions for the DSA efficiency by comparing the $\gamma$-ray flux from
simulated clusters with the Fermi-LAT upper-limit flux levels of observed clusters.
Even with the relatively less efficient model based on the hybrid simulation results of \cite{caprioli14},
in which $\eta\approx 0.05$ for $M_A= 5$ quasi-parallel shocks, and the consideration of the random magnetic field directions,
they find that about 10-20 \% of simulated clusters have the predicted $\gamma$-ray flux levels above the Fermi-LAT upper limits. 
So the authors suggested that only if $\eta\le 10^{-3}$ for all Mach number shocks, which results in the average value of 
$\langle X_{\rm CR}\rangle\lesssim 0.01$ in the ICM,
the predicted $\gamma$-ray fluxes from simulated clusters can stay below the Fermi-LAT upper limits.
This agrees with the conclusion of \cite{fermi14}, which predicted $\langle X_{\rm CR}\rangle\lesssim 0.0125-0.014$ 
based on the analysis of four year Fermi-LAT data.

Non-detection of $\gamma$-ray emission from galaxy clusters might be explained, if the CR proton acceleration 
is much less efficient than expected in the current DSA theory (i.e.~$\eta\lesssim 10^{-3}$ for $M_s\sim 3$). 
In that regard, the proton acceleration at weak shocks in the low density, high beta ICM plasma needs to be investigated further,
since so far most of hybrid/PIC plasma simulations have focused on strong shocks in $\beta_p\lesssim 1$ ISM and solar wind plasma.

Finally, armed with our new understandings based on the recent plasma hybrid simulations \cite{caprioli14, caprioli14B},
it is worth examining if strong accretions shock can accelerate CR protons to ultra-high energies.
The protons are expected to be accelerated efficiently via DSA only 
in the quasi-parallel portion of the outermost surfaces encompassed with accretion shocks.
There magnetic fields could be amplified via Bell's non-resonant hybrid instability 
by a factor of $B/B_0\propto \sqrt{M_A}$ \cite{caprioli14B}, where $B_0\sim 0.01 \mu {\rm G}$ and $M_A\sim 300$.
So it is reasonable to assume the magnetic field strength at external accretion shocks is $B\sim 0.1 \mu {\rm G}$,
about one order of magnitude smaller than that typically adopted in the previous studies \cite[e.g.,][]{norman95,kang96}.
Considering the photo-pair energy losses, protons can be accelerated up to $E_{\rm p,max}\sim 10^{18}$~eV at
quasi-parallel accretion shocks (see Figure 1).

\section{Summary}

\begin{enumerate}

\item Astrophysical plasmas consist of both thermal and CR particles that are closely coupled with
permeating magnetic fields and underlying turbulent flows. So understanding the complex network of physical interactions among these components,
especially in the high beta collisionless ICM plasma, 
is crucial to the study of the particle acceleration at structure formation shocks (see Figure 6).

\item Gravitational energy associated with hierarchical clustering of the large-scale-structures must be 
dissipated at structure formation shocks 
into several different forms: heat, CRs, turbulence and magnetic fields \cite{ryu03}.

\item The vorticity generated by curved shocks decays into turbulence behind the shock, which in turn cascades into MHD/plasma waves in
a wide range of scales and amplify magnetic field via turbulence dynamo \cite{ryu08}.

\item  There is growing observational evidence indicating the presence of weak shocks, relativistic electrons,
microgauss level magnetic fields, and turbulence in the ICM of galaxy clusters \cite{brunetti14}.

\item CR protons are expected to be accelerated mainly at quasi-parallel shocks. 
For weak internal shocks ($M_s\lesssim 3$) with high kinetic energy fluxes that form in the ICM, 
the CR proton acceleration efficiency is likely to be $\eta< 0.01$ in order to explain the non-detection of
$\gamma$-ray emission from galaxy clusters due to inelastic {\it p-p} collisions in the ICM \cite{vazza16CR}.

\item At quasi-parallel portion of strong external accretion shocks, CR protons could be accelerated to $\sim 10^{18}$~eV,
if the preshock magnetic fields can be amplified to $\sim 0.1\mu {\rm G}$ via CR streaming instabilities \cite{kang96,caprioli14}. 

\item CR electrons are expected to be accelerated preferentially at quasi-perpendicular shocks \cite{guo14}.
Radio relics detected in the outskirts of merging clusters seem to reveal radiative signatures 
of relativistic electrons accelerated at merger-driven shocks mostly with $M_s\sim 2-3$ \cite{kang12}.     

\item  The injection of protons and electrons from thermal or suprathermal populations to the
DSA process at collisionless shocks involves plasma kinetic processes such as excitation of waves
by micro-instabilities as well as shock drift acceleration and shock surfing acceleration \cite{kang14}.
During the last decade significant progress been made in that front through PIC/hybrid plasma simulations 
of non-relativistic shocks \cite{caprioli14, guo14}.

\end{enumerate}

\section{Acknowledgements}
This work was supported by the National Research Foundation of Korea through grants NRF-2014R1A1A2057940 and NRF-2016R1A5A1013277.
The author would like to thank D. Ryu for helpful comments on the paper.



%% References

\nocite{*}
\bibliographystyle{elsarticle-num}
\bibliography{HKang}






% \end{thebibliography}

\end{document}